\documentclass[conference]{IEEEtran}
\IEEEoverridecommandlockouts

\usepackage{booktabs}
\usepackage{xspace}
\usepackage{colortbl}
\usepackage{multirow}

\usepackage{cite}
\usepackage{amsmath,amssymb,amsfonts}
\usepackage{algorithmic}
\usepackage{graphicx}
\usepackage{textcomp}
\usepackage{xcolor}
\def\BibTeX{{\rm B\kern-.05em{\sc i\kern-.025em b}\kern-.08em
    T\kern-.1667em\lower.7ex\hbox{E}\kern-.125emX}}
\begin{document}

\definecolor{mygray}{rgb}{0.95, 0.95, 0.95}
\definecolor{myblue}{rgb}{0.41, 0.50, 0.57}
\definecolor{greyblue}{RGB}{177,221,240}
\definecolor{lightgold}{RGB}{249,247,237}
\definecolor{peacockblue}{RGB}{27,161,226}

\NewDocumentCommand\emojititle{}{
    $\vcenter{\hbox{\includegraphics[height=1.2em]{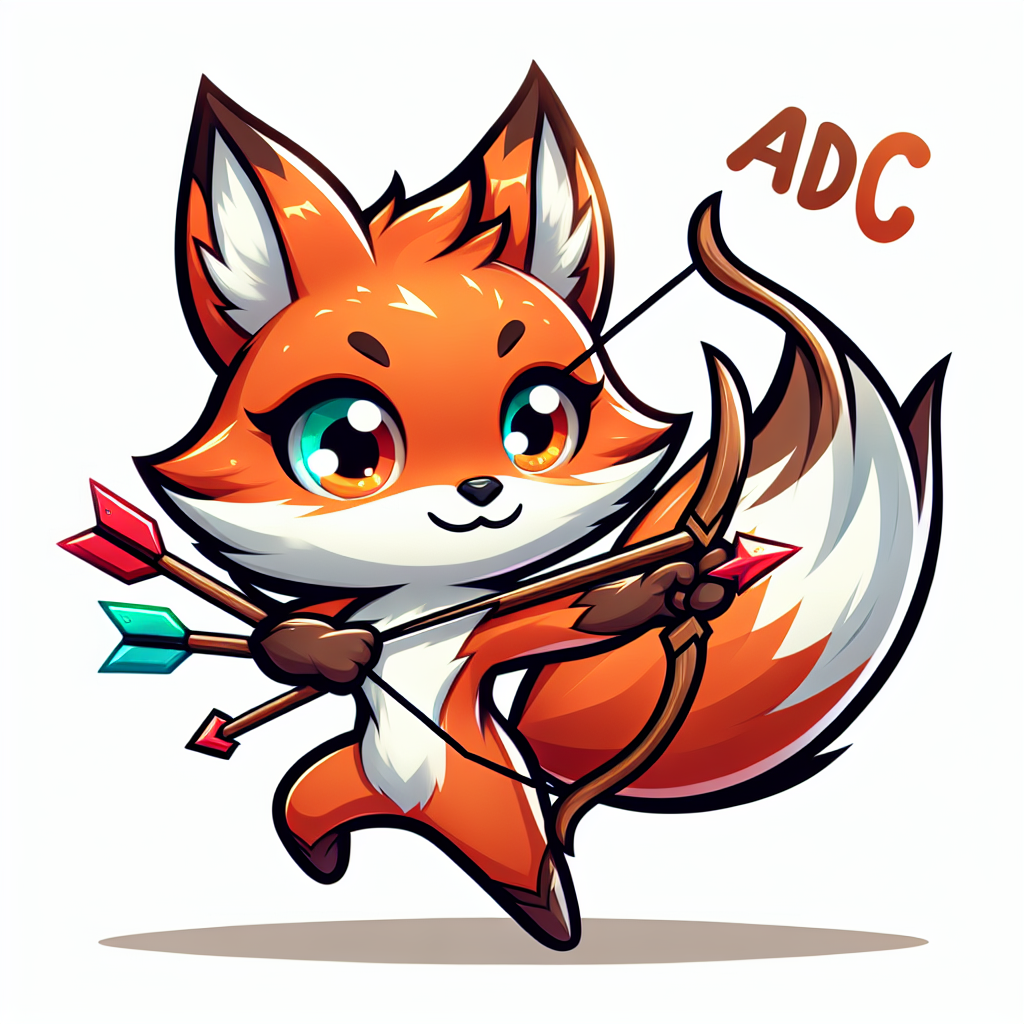}}}$
}

\newcommand{\model}{\textsc{ADC}}

\title{\emojititle{}\model{}: Enhancing Function Calling Via Adversarial Datasets and Code Line-Level Feedback\thanks{\\$^{\S}$Equal Contribution.\\$^{\star}$This research was primarily done during the internship at Alibaba Group.\\$^{\dagger}$Corresponding authors.}}

\author{
\IEEEauthorblockN{
Wei Zhang$^{12\star\S}$, Yi Zhang$^{1\S}$, Li Zhu$^{2}$, Qianghuai Jia$^{2\dagger}$, Feijun Jiang$^{2\dagger}$, Hongcheng Guo$^{1}$, \\ Zhoujun Li$^{1\dagger}$, Mengping Zhou$^{2}$
}
\IEEEauthorblockA{
\textit{$^{1}$CCSE, Beihang University.}
\textit{$^{2}$Alibaba International Digital Commerce.} \\
\texttt{zwpride@buaa.edu.cn; zhangyi2021@buaa.edu.cn; qianghuai.jqh@alibaba-inc.com;}
}
}


\maketitle
\begin{abstract}
Large Language Models (LLMs) have made significant strides in Natural Language Processing and coding, yet they struggle with robustness and accuracy in complex function calls. To tackle these challenges, this paper introduces ADC, an innovative approach that enhances LLMs' ability to follow function formats and match complex parameters. ADC utilizes a high-quality code fine-tuning dataset with line-level execution feedback, providing granular process supervision that fosters strong logical reasoning and adherence to function formats. It also employs an adversarial dataset generation process to improve parameter matching. The staged training methodology capitalizes on both enriched code datasets and refined adversarial datasets, leading to marked improvements in function calling capabilities on the Berkeley Function-Calling Leaderboard (BFCL) Benchmark. The innovation of ADC lies in its strategic combination of process supervision, adversarial refinement, and incremental learning, setting a new standard for LLM proficiency in complex function calling.
\end{abstract}


\section{Introduction}
Large Language Models (LLMs) including ChatGPT, GPT-4, and Gemini have achieved unparalleled proficiency in Natural Language Processing (NLP) and coding due to extensive pre-training, enhancing their problem-solving and instruction-following abilities. Function calling enables LLMs to effectively utilize external tools and APIs~\cite{zhang2024lemurlogparsingentropy,zhang2024eclipsesemanticentropylcscrosslingual}, enhancing their capabilities beyond basic text generation. This makes interactions more powerful by allowing them to perform specific operations as needed.

CodeAlpaca~\cite{codealpaca} uses 21 seed tasks and generates a 20k dataset via self-instruct~\cite{wang2022self}, while Wizardcoder~\cite{luo2023wizardcoder}, MagicCoder~\cite{wei2023magicoder}, and WaveCoder~\cite{yu2024wavecoderwidespreadversatileenhancement}
apply advanced heuristics and novel data generation processes based on open-source code snippets and code instruction data to enhance the complexity of initial code instructions. ESETR~\cite{moon2024efficientscalableestimationtool} retrieves the most relevant tools for a given query, xLAM~\cite{zhang2024xlamfamilylargeaction} uses mixture-of-expert architectures and dataset pipeline and Granite-FC-Model~\cite{abdelaziz2024granitefunctioncallingmodelintroducing}
trains with a multi-task training approach on seven fundamental tasks encompassed in function calling~\cite{zhang2024mabcmultiagentblockchaininspiredcollaboration}.
Despite these efforts, existing methods lack robustness and accuracy, especially with complex function calls and diverse programming scenarios. These models often \textbf{fail to follow function formats}, necessitating strong logical reasoning to apply best practices effectively.
Additionally, there remains \textbf{a significant gap in complex parameter matching}, affecting the efficiency and accuracy of code generation.

To address these challenges, we introduce a comprehensive approach, enhancing function calling via \textbf{A}dversarial \textbf{D}atasets and \textbf{C}ode line-level feedback (\model{}), that uses code with execution feedback for process supervision, enhancing logical reasoning and function format following ability.
It also employs adversarial function call datasets to improve parameter matching.
Staged training driven by them boosts LLM function calling. Specifically, we meticulously construct a high-quality code fine-tuning dataset with code samples annotated with line-level execution feedback.
These annotations are crucial, as they provide granular process supervision, enabling \model{} to learn from detailed, real-world code execution scenarios for stronger logical reasoning and better function format following ability.
To further boost the robustness and accuracy of parameter matching, we employ an adversarial process in which an LLM generator creates challenging function calling data and an LLM discriminator evaluates them, where each strives to outsmart the other and thereby refine the datasets.
Our staged training process strategically leverages both the enriched code dataset and the refined function calling dataset. This approach allows \model{} to progressively build and refine its understanding of function calling, leading to substantial performance gains. We achieve superior results when tested with the Berkeley Function-Calling Leaderboard ~\cite{berkeley-function-calling-leaderboard} (BFCL) Benchmark. Our results demonstrate significant improvements over other strong baselines, underscoring the efficacy of our comprehensive training regimen and dataset refinement techniques.
Overall, our main contributions are:\begin{itemize}
\item \textbf{Line-Level Code Execution Feedback}: We create a detailed code dataset with line-level execution feedback, enabling \model{} to learn precise logical reasoning and function format following for function calling from real-world code scenarios under process supervision.
\item \textbf{Adversarial Process For Function Calling}: We implement an adversarial process that generates challenging function calling data, enhancing \model{}'s ability to handle diverse and complex parameter matches.
\item \textbf{Staged Training Process}: Our structured staged training leverages enriched and refined datasets, significantly improving function calling accuracy on BFCL.
\end{itemize}

\begin{figure*}[h]
\centering
\includegraphics[width=1.0\textwidth]{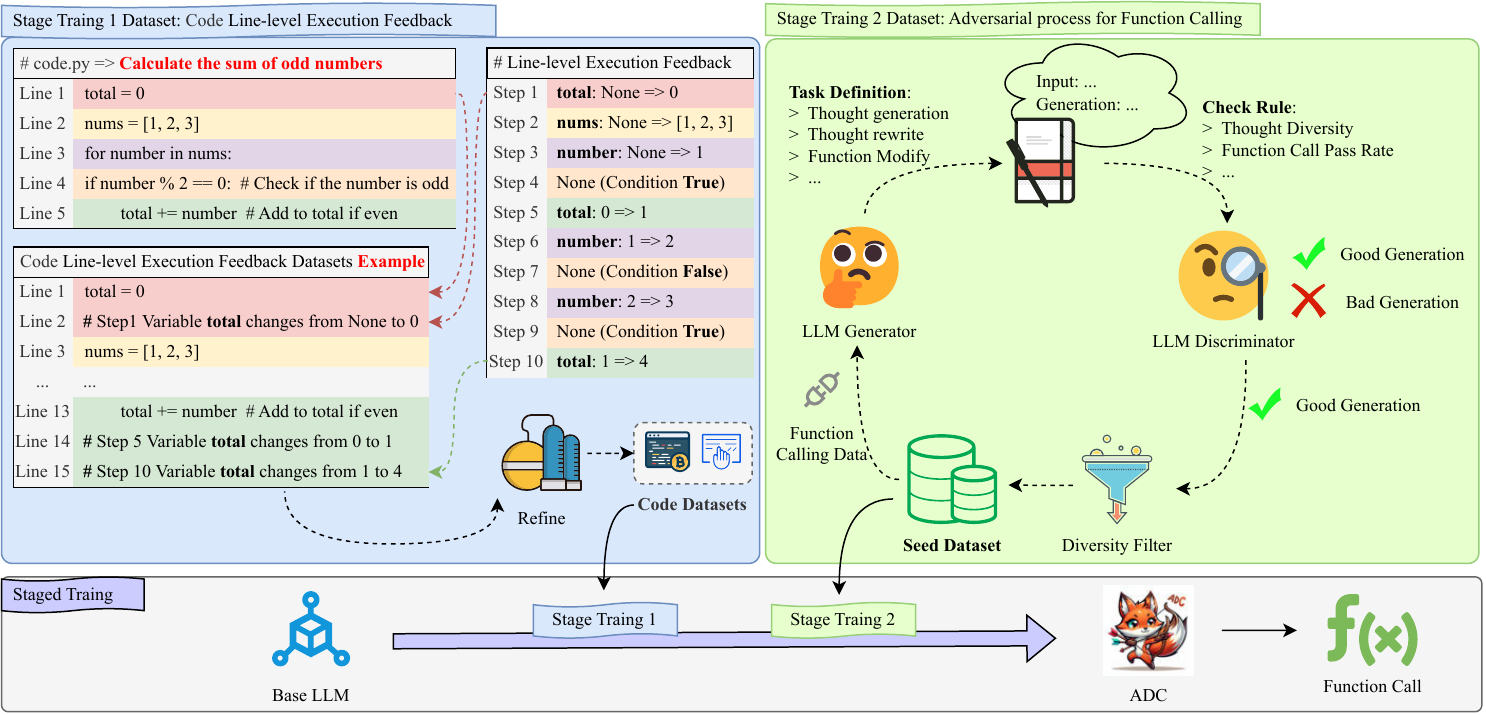}
\caption{Overview of \model{}. We first create a detailed code dataset with line-level execution feedback by executing the code and embedding the feedback into the code. Then, we employ an LLM generator and an LLM discriminator to refine the function calling dataset. The staged training process leverages both datasets to improve the function calling ability of \model{}.}
\label{fig:adc}
\vspace{-10pt}
\end{figure*}

\section{Methodology}
In Figure~\ref{fig:adc}, we provide a comprehensive overview of \model{}. We begin by detailing code line-level execution feedback followed by the adversarial process for function calling and conclude with an introduction to our staged training process.

\begin{table*}[t]
\centering
    \caption{\small Performance comparison on the BFCL-v2 leaderboard.}
    \label{table:comparison}
    \resizebox{0.8\linewidth}{!}{
        \begin{tabular}{ccccccc}
            \hline
            Model & Overall$\uparrow$ & AST$\uparrow$  & Execution$\uparrow$ & Irrelevance$\uparrow$ & Relevance$\uparrow$ \\
            \hline
            Meta-Llama-3-70B-Instruct~\cite{meta2024llama3} & 81.59 & 80.15 & 88.04 & 50.47 & 92.68 \\
            GPT-4-0125-Preview~\cite{achiam2023gpt} & 85.79 & 85.5 & 89.25 & 61.35 & 97.56 \\
            GPT-4o-mini-2024-07-18~\cite{achiam2023gpt} & 83.35 & 80.51 & 87.95 & 79.20 & 80.49 \\
            Claude-3-Opus-20240229~\cite{claudeai2024} & 80.88  & 79.42 & 87.39 & 56.15 & 85.37 \\
            GPT-3.5-Turbo-0125~\cite{brown2020languagemodelsfewshotlearners} & 66.19  & 60.14 & 65.88 & 69.97 & 87.8 \\
            \hline
            Meta-Llama-3-8B-Instruct~\cite{meta2024llama3} & 62.70 & 58.89 & 72.62 & 22.88 & 78.05 \\
            Meta-Llama3.1-8B-Instruct~\cite{{meta2024llama3}} & 63.19 & 57.43 & 76.39 & 13.72 & 82.93 \\
            Hermes-2-Pro-Llama-3-8B~\cite{Hermes-2-Pro-Llama-3-8B} & 66.18   & 64.18 & 74.05 & 55.16 & 53.66 \\
            Hermes-2-Pro-Mistral-7B~\cite{Hermes-2-Pro-Mistral-7B} & 65.44  & 60.82 & 74.25 & 38.55 & 75.61 \\
            Hermes-2-Theta-Llama-3-8B~\cite{Hermes-2-Theta-Llama-3-8B} & 64.83  & 61.08 & 72.54 & 62.66 & 51.22 \\
            Functionary-Small-v3.1~\cite{functionary-small-v3.1} & \textbf{80.21} & \textbf{78.64} & 83.45 & 68.36 & 85.37 \\
            Functionary-Small-v3.2~\cite{functionary-small-v3.2} & 78.96 & 76.16 & 83.04 & 72.32 & 80.49 \\
            Gorilla-OpenFunctions-v2~\cite{gorilla-openfunctions-v2} & 79.10 & 73.17 & 84.96 & 73.13 & \textbf{85.37} \\
            xLAM-7b-fc-r\cite{zhang2024xlamfamilylargeaction} & 79.41 & 72.77& 85.68 & \textbf{79.76} & 80.49 \\
            \hline
            \textbf{\model{} (Our Method)} & 79.01 & 70.46 & \textbf{87.50} & 75.67 & 82.89 \\ 
            \hline
        \end{tabular}
}
\vspace{-15pt}
\end{table*}

\subsection{Code Line-Level Execution Feedback}
We use code with line-level execution feedback as process supervision to enhance the logical reasoning ability of LLM to improve the function format following, where the code data is sampled from publicly available datasets, and the line-level execution feedback is achieved by executing the code to track variable changes.

\subsubsection{Data Collection}
We gather source code from two key datasets: CodeNet~\cite{puri2021codenet}, which comprises approximately 14 million code snippets, and POJ104~\cite{poj104}, a smaller dataset consisting of 52,000 code snippets focused on 104 algorithmic problems. CodeNet serves as the primary source, while POJ104 is incorporated to enhance its diversity.

\subsubsection{Feedback Generation}
As depicted in Figure~\ref{fig:adc}, line-level execution feedback consists of incremental variable changes observed during the line execution. We format the feedback as follows: \texttt{v: x => y at line i}, where \texttt{v} means the variables that change when code execution, \texttt{x} and \texttt{y} means the values from which and to which the variables change, and \texttt{i} means the specific line of code execution.

For Python snippets, we utilize a package called \texttt{pysnooper} to generate the feedback. We design a wrapper tracing program that executes code and retrieves the corresponding variable changes as line-level execution feedback. The execution is performed using Python version 3.10, with the AIZU Online Judge~\cite{aizu} and AtCoder ~\cite{atcoder} packages installed in the Python environment.

For C and C++ snippets, we modify the original code using regex-based rules to print variable changes to the standard error stream. The snippets are compiled using the g++ 11.4.0 compiler with C++11 language features enabled. To address common defects found in many C and C++ code snippets, we implement by including the GNU C++ header \texttt{<bits/stdc++.h>} to encompass all standard library headers and insert the statement \texttt{using namespace std;}.

\subsubsection{Data Refine}
Following the execution of the code snippets and the collection of execution feedback, we refine the collected data.

\paragraph{Redundancy Reduction}Due to the presence of program structures such as loops, recursion, and searches, the execution feedback can become exceedingly lengthy, occasionally exceeding 1 million characters for a single code snippet. Thus, we develop post-processing techniques aimed at reducing redundancy in terms of information entropy. This includes limiting the number of generated feedback lines by replacing intermediate steps of each line-variable pair with ellipses `...`, and imposing a limit of 10 steps per pair.

\paragraph{Length-based Filtering}Statistical analysis indicates that 90\% of code snippets fall below 1000 characters in length, and the combined length of code and feedback typically remains below 2000 characters. Consequently, we filter out code snippets whose combined length with feedback exceeds 2048 characters.

\paragraph{Invalid Code Filtering}While code snippets that fail to execute successfully or do not produce correct outputs can provide insights for code composition, we prioritize including correct solutions to enhance the problem-solving capabilities of LLMs. Therefore, we exclude all code snippets that fail to terminate properly, encounter runtime errors, or yield incorrect results during execution.

\paragraph{Non-informative Filtering}Even when execution is successful, some code snippets do not generate any execution feedback due to the absence of intermediate variables. These snippets are often simplistic, implementing basic functions such as `A + B`. Since they do not contribute to the LLM's comprehension of execution context, we filter out such non-informative data.

\subsubsection{Feedback Embedding}
We obtain a dataset comprising both code and feedback and add a prefix to each feedback line to denote the step number, thereby elucidating program structures like \texttt{Step STEP-NUMBER, Variable VARIABLE-NAME changes from OLD-VALUE to NEW-VALUE.} We embed the feedback lines directly under their corresponding code lines to facilitate model training.

Additionally, we annotate the code with input data and standard output. To ensure consistency with the code syntax and facilitate comprehension by the LLM, we format the feedback, input data, and standard output as comments, utilizing language-specific comment symbols. Figure~\ref{fig:adc} depicts the code snippet with embedded feedback.

\subsection{Adversarial process for Function Calling}
\subsubsection{Seed Dataset Collect}
To further enhance the robustness and accuracy of the function call predictions, we employ an adversarial process involving an LLM generator and an LLM discriminator. This process is bootstrapped using two seed datasets: 60k xlam-function-calling~\cite{liu2024apigen} from Salesforce and ToolBench~\cite{guo2024stabletoolbench} from OpenBMB. These foundational datasets provide a rich and diverse collection of function call scenarios that serve as the starting point for our refinement process.

\subsubsection{Criteria for Data Evaluation}
The criteria for data evaluation focus on whether the generated thought (logic) behind the function call is reasonable and optimized. It also includes assessing the potential complexity and variety of function call chains, such as parallel, chain-like, and network-like calls. A well-reasoned thought must consider both single-turn and multi-turn (chain/network) function calls to ensure the generated data accurately reflects the diversity and complexity of real-world scenarios.

\subsubsection{LLM Generator}
LLM generator, guided by the evaluation criteria and current seed data, introduces variations, edge cases, and complex interactions to generate more intricate and valuable scenarios. By incorporating these changes, the data generator creates new function call sequences and complex function parameters that simulate a wide range of real-world complexities and edge cases. 

\subsubsection{LLM Discriminator}
LLM discriminator evaluates the data generated by the generator based on the predefined criteria. Data that meet the standards are incorporated into the seed dataset, while those that do not are dropped. LLM discriminator ensures that the generated data not only exhibits complexity and value on function calling sequences and parameters but also maintains realism and logical consistency, thus enhancing the overall quality of the dataset.

\subsubsection{Iterative Adversarial Process}
Our uses an iterative adversarial process for continuous improvement. Starting with seed datasets, the LLM generator introduces variations, and the LLM discriminator ensures quality. This dynamic interaction covers single-turn and multi-turn calls, making the dataset comprehensive and challenging.

\subsection{Staged Train Process}
We believe that a high parameter matching accuracy is meaningful only when the format is highly accurate; otherwise, it won't improve overall function call effectiveness. Due to the limited availability of high-quality code data for execution feedback, we opt for phased training instead of mixed training to better emphasize code data.

We fine-tune base LLM with code dataset with line-level execution feedback to enhance the function format following. Then we continue to fine-tune with adversarial function call datasets for better parameter matches in challenging scenarios.

\section{Experiments}

\subsection{Experimental Setup}
We have implemented \model{} on CentOS 7, powered by two Intel(R) Xeon(R) Platinum 8480+ with a total of 112 cores, 8 * NVIDIA H800 (80G), and 528 GB of memory. The software setup includes NVIDIA-SMI version 535.129.03 and CUDA version 12.2. We choose Llama3.1-8B-Instruct as our base to evaluate \model{}.

\subsection{Benchmark and Metrics}
The Berkeley Function-Calling Leaderboard (BFCL) Benchmark~\cite{berkeley-function-calling-leaderboard} provides a comprehensive evaluation framework for assessing an agent's capability to reason about and execute function calls across various programming languages and domains. 
With over 2,200 test cases in Java, JavaScript, and Python, the benchmark measures Abstract Syntax Tree (AST) accuracy, executable accuracy, irrelevance, and relevance detection. 
Our evaluation utilizes the latest BFCL v2, which introduces live function calls and real-world user-contributed scenarios to address data contamination, bias, and fairness. The v2 dataset reflects real-world distributions more accurately, with a higher demand for selecting among multiple functions and a reduced demand for parallel calls.

Following BFCL~\cite{berkeley-function-calling-leaderboard}, We evaluate \model{} on five metrics. \textbf{AST Summary} compares the function's abstract syntax tree to the ground truth to check for correctness in calls, parameters, and types. \textbf{Execution Summary} evaluates the output of generated and ground-truth function calls for both REST and non-REST APIs. \textbf{Irrelevance} and \textbf{Relevance} detect and assess the model's ability to avoid irrelevant function calls and recognize correct function calls. \textbf{Overall Accuracy} is the weighted average of all data splits.

\subsection{Main Results}
As detailed in Table \ref{table:comparison} on the BFCL-v2 leaderboard, includes three categories of models: Mainstream large-scale commercial models like the GPT ~\cite{brown2020languagemodelsfewshotlearners} series, Meta-Llama~\cite{meta2024llama3}, and Claude~\cite{claudeai2024}; open source models comparable in size to our own; and \model{}.

\model{} sets a new benchmark in the Execution metric with an outstanding score of 87.50, showcasing its superior ability to generate function calls that are not only syntactically correct but also logically coherent. This performance highlights \model{} proficiency in logic reasoning, function format following, and parameter match, a focal point of our methodological improvements. Although \model{} achieved a notable Overall Score of 79.01, it displayed a consistently strong performance across all metrics. Its execution strength is particularly remarkable when compared to competitors like xLAM-7b-fc-r\cite{zhang2024xlamfamilylargeaction}, which narrowly leads in the overall score. The strategic emphasis on execution and logical reasoning in the development of our model sometimes results in concessions in AST and relevance scores for better execution. With a relevance score of 82.89, \model{} may not have reached the peak, but it still confirms our model's effectiveness in generating contextually appropriate function calls that align well with the provided prompts. 

In summary, \model{} excels in execution, reflects a strong overall performance, and adeptly balances logical coherence with relevance, establishing it as a formidable contender in its class.

\subsection{Ablation Study}

\begin{table}[t]
\centering
    \caption{\small Ablation Study for \model{}.}
    \label{table:abalation}
    \resizebox{0.9\linewidth}{!}{
        \begin{tabular}{ccc}
            \hline
            Model & Method & Overall$\uparrow$ \\
            \textbf{\model{}} & - & 79.01 \\ 
            \textbf{\model{}} & w/o code dataset & 67.73 \\ 
            \textbf{\model{}} & w/o function calling dataset & 55.78 \\ 
            \textbf{\model{}} & prefix embed code feedback & 59.89 \\ 
            \textbf{\model{}} & suffix embed code feedback & 65.13 \\
            \hline
        \end{tabular}
}
\end{table}

We conducted an ablation analysis on \model{} to dissect the contribution of each component. This involved examining code dataset and function calling dataset, alongside distinct embedding techniques, namely prefix embedding and suffix embedding. These embedding methods diverge from the standard approach by incorporating execution feedback directly into the code concatenation process, rather than employing row-level embedding.

As detailed in Table \ref{table:abalation}, the absence of the code dataset led to a notable decrease in performance, with the overall score dropping to 67.73. This highlights its crucial contribution to \model{}'s ability to follow function format. A more pronounced decline was observed when the function calling dataset was excluded, plummeting the performance to 55.78. This underscores its significance in facilitating complex parameter matching, a vital task for the model. Results of embedding techniques show more apparent differences. Employing either prefix or suffix embedding to incorporate code feedback resulted in diminished scores of 59.89 and 65.13, respectively. This outcome strongly suggests that our unique approach of line-level embedding, which embeds context-rich feedback directly associated with each line of code, is significantly more effective in enhancing the model's proficiency in function calling.

Overall, the ablation study highlights the indispensable nature of both datasets and the efficacy of line-level embedding in \model{}, setting a robust foundation for function format following and parameter matching.

\section{Conclusion}

In conclusion, our proposed approach, \model{}, effectively enhances function calling accuracy in Large Language Models through detailed line-level execution feedback, an adversarial data generation process, and a structured staged training regimen. These innovations address critical shortcomings related to function format following, and better parameter match, leading to significant performance improvements on the Berkeley Function-Calling Leaderboard. Our findings underscore the potential of refined training methodologies in advancing LLM capabilities.

\clearpage
\bibliographystyle{IEEEtran}
\bibliography{IEEEabrv,custom}

\end{document}